\documentstyle[12pt]{article}
\catcode`\@=11 \@addtoreset{equation}{section}\catcode`\@=12
\newcommand{\beq}[1]{\begin{equation}\label{#1}}
\newcommand{\eeq}{\end{equation}}
\newcommand{\bear}[1]{\begin{eqnarray}\label{#1}}
\newcommand{\ear}{\end{eqnarray}}

\newcommand{\rf}[1]{(\ref{#1})}
\newcommand{\R}{\mbox{\rm I$\!$R} }
\newcommand{\M}{\mbox{\rm I$\!$M} }
\newcommand{\foom}[1]{\protect\footnotemark[#1]}
\newcommand{\email}[2]{\footnotetext[#1]{e-mail: #2}}%
\newcommand{\Title}[1]{\noindent {\Large #1} \\}

\newcommand{\eps}{\varepsilon }

\begin{document}
\begin{center}
\Title{\large\bf Multicomponent perfect fluid with
variable parameters in $n$ Ricci-flat spaces}

\bigskip

\noindent{\normalsize\bf
J.-M. Alimi\foom 1,
V. R. Gavrilov\foom 2 and
V. N. Melnikov\foom 3}

\medskip

\noindent{\it LUTH, Observatoire Paris-Meudon, \\ Bat. 18,   Pl. J. Janssen,
92195 MEUDON, France \\ and \\

Centre for Gravitation and Fundamental Metrology
\\
VNIIMS, 3-1 M. Ulyanovoy St., Moscow 119313, Russia}

\vspace{5mm}
{\bf Abstract}
\end{center}

$D$-dimensional cosmological model describing the evolution of a
multicomponent perfect fluid with variable barotropic parameters
in $n$ Ricci-flat spaces is investigated. The equations of motion
are integrated for the case, when each component possesses an
isotropic pressure with respect to all  spaces. Exact solutions
are presented in the Kasner-like form.
 Some explicit examples are given:
$4$-dimensional model with an isotropic accelerated expansion
at late times and  $(4+d)$-dimensional model describing a compactification
of extra dimensions.
\email 1 {jean-michel.alimi@obspm.fr}
\email 2 {gavr@rgs.phys.msu.su}
\email 3 {melnikov@rgs.phys.msu.su;}


\section{Introduction}

The necessity of studying multidimensional models of gravitation and
cosmology \cite{Mel2,Mel,Mel02} is motivated by several reasons. First, the
main trend of modern physics is the unification of all known fundamental
physical interactions: electromagnetic, weak, strong and gravitational
ones. During the recent decades there has been a significant progress in
unifying weak and electromagnetic interactions, some more modest
achievements in GUT, supersymmetric, string and superstring theories.

Now, theories with membranes, $p$-branes and more vague M- and F-theories
are being created and studied. Having no definite successful theory of
unification now, it is desirable to study the common features of these
theories and their applications to solving basic problems of modern
gravity and cosmology.  Moreover, if we really believe in unified
theories, the early stages of the Universe evolution and black hole physics,
as unique superhigh energy regions, are the most proper and natural arena
for them.

Second, multidimensional gravitational models, as well as scalar-tensor
theories of gravity, are theoretical frameworks for describing
possible temporal and range variations of fundamental physical constants
\cite{3,4,5,6}.

Lastly, applying multidimensional gravitational models to basic problems
of modern cosmology and black hole physics, we hope to find answers to such
long-standing problems as : singular or nonsingular initial states, creation
of the Universe, its flatness, creation of matter and its entropy, acceleration,
coincidence and
cosmological constant, origin of inflation and specific scalar fields
which may be necessary for its realization, isotropization and graceful
exit problems, stability and nature of fundamental constants \cite{4},
possible number of extra dimensions, their stable compactification etc.

The discovery of the accelerated expansion of the Universe and the
fact that the flat Friedmann model with the cosmological constant
or quintessence now fits best the set of different observational
data, created the problems of dark matter and dark energy. This is
a real revolution in modern physics as we do not know now what
really the dark matter (0.20 of 0.30) and what is the dark energy (0.70)
of the total energy are. Even attempts to explain it via the
cosmological constant or quintessence seem to change one puzzle
with another one as necessary vacuum properties or exotic scalar
fields with or without strange potentials are still waiting to
find their place in rigorous theories, not speaking about their
real experimental confirmation.

In \cite{GMX} we showed that the cosmic acceleration and
coincidence problems may be solved by using an x-fluid as a
quintessence and a viscous fluid as a normal matter. Viscosity of
the normal matter can be explained by its own multicomponent
structure. As is known, a mixture of different fluids admits a
description as a single viscous fluid.  We adopted the "second
equations of state" in the form of some special metric dependence
of the bulk and shear viscosity coefficients. These "second
equations of state" generalize the so called "linear dissipative
regime" in FRW world model, when the bulk viscosity coefficient is
linearly proportional to the Hubble parameter. We studied
$D$-dimensional homogeneous anisotropic cosmology, which allows to
describe the dynamical compactification of the extra dimensions
(see, for instance, our paper \cite{Gavrilov} on viscous
cosmology). Other 2-component models in many dimensions also
having the acceleration were found: with the cosmological constant
in \cite{a2}, with a perfect fluid in \cite{a1}, with 2
non-Ricci-flat spaces \cite{a3}, with p-branes and static internal
spaces in \cite{a4}, with scalar fields having exponential
potentials in \cite{a8} and in four dimensions with a perfect
fluid and a scalar field with the exponential potential in
\cite{a6,a7} using methods developed in our multidimensional
approach.

Here with the same aim the $D$-dimensional cosmological model describing the evolution of a
multicomponent perfect fluid with variable barotropic parameters
in $n$ Ricci-flat spaces is investigated. The equations of motion
are integrated for the case, when each component possesses an
isotropic pressure with respect to  all spaces. The exact
solutions are presented in the Kasner-like form.
Some examples are
given. The first example is $4$-dimensional model with the Kasner-like
behavior near the initial singularity and
an isotropic accelerated expansion at late times.
The other example is $(4+d)$-dimensional model describing a contraction
of the internal space accompanied by the expansion of  the external
space at early times.

\section{The  model}

Following  papers \cite{92}-\cite{96}
we consider the metric
\bear{metr}
{\rm d}s^2=-{\rm
e}^{2\gamma(t)}{\rm d}t^2  + \sum_{i=1}^{n}\exp[2x^{i}(t)]{\rm
d}s_i^2,
\ear
on  $D$-dimensional space-time manifold
 \beq{manifold}
 \M=\R\times
M_1\times\ldots\times M_n,
\eeq
 where ${\rm d}s_i^2$ is a metric
of the Ricci-flat factor  space $M_{i}$ of dimension $d_i$,
$\gamma(t)$ and $x^{i}(t)$ are scalar functions of the cosmic time
$t$, $a_i\equiv\exp[x^{i}]$ is the scale factor of the space $M_i$
and the function  $\gamma(t)$ determines a time gauge. The
synchronous time $t_s$ is defined by the equation ${\rm
d}t_s=\exp[\gamma(t)]{\rm d}t$.

Under this assumption the Ricci tensor for the metric \rf{metr}
has the following non-zero components
\begin{eqnarray*}
R_0^0&=& {\rm e}^{-2\gamma}\left( \sum_{i=1}^{n}
d_{i}(\dot{x}^{i})^2+  \ddot{\gamma_0}-
\dot{\gamma}\dot{\gamma_0}\right)
\\
R_{n_i}^{m_i}&=& {\rm e}^{-2\gamma} \left[\ddot{x}^{i}+
\dot{x}^{i}(\dot{\gamma_0}-\dot{\gamma})\right] \delta_{n_i}^{m_i}
\end{eqnarray*}
with the definition
\beq{gamma0}
\gamma_0=\sum_{i=1}^{n} d_{i}x^{i}.
\eeq
Here indices $m_i$ and $n_i$  for $i=1,\ldots,n$ run from
($D-\sum_{j=i}^{n}d_j$) to ($\sum_{j=i}^i d_j$), where
\linebreak
$D=1+\sum_{i=1}^{n}d_i={\rm dim}\M $.

We consider a source of
gravitational field in the form of multicomponent perfect fluid.
In comoving coordinates the energy-momentum tensor of such a source
reads
\bear{tensorsum}
&& T^{M}_{N} = \sum_{s =1}^{m} T^{M(s)}_{N}, \\
\label{tensor}
&&(T^{M (s)}_{N})={\rm diag}\left(-{\rho^{(s)}}(t),
\underbrace{p_{1}^{(s)}(t),\ldots , p_{1}^{(s)}(t)}_{d_1 times},\ldots ,
\underbrace{p_{n}^{(s)}(t),\ldots , p_{n}^{(s)}(t)}_{d_n times}\right)
\quad ,
\ear
Furthermore,
we suppose that
the barotropic equation of state
for the perfect fluid components
is given by
\beq{pressure}
{p_{i}^{(s)}}(t) =\left(1-h_{i}^{(s)}(x)\right){\rho^{(s)}}(t),\ \
s=1,\ldots,m,
\eeq
where  variable barotropic parameters are given by
\beq{parameter}
h_{i}^{(s)}(x)=\frac{1}{d_i}\frac{\partial}{\partial x^i}\Phi^{(s)}(x),
\ \
i=1,\ldots,n
\eeq
with  an arbitrary smooth function $\Phi^{(s)}(x)$ on $\R^n$.

The equation of motion
$\bigtriangledown_{M} T^{M (s)}_{0}=0$ for the perfect fluid component
described by the tensor \rf{tensor} reads
\beq{conservation}
\dot{\rho}^{(s)}+
\sum_{i=1}^{n}d_{i}\dot{x}^{i}\left(\rho^{(s)} + p_{i}^{(s)}\right)=0.
\eeq
After using  equations of state \rf{pressure} via \rf{conservation},
integrals of motion may be obtained in the form
\beq{integrals}
A^{(s)}={\rho^{(s)}}
\exp\left[2\gamma_{0} - \Phi^{(s)}(x)\right],\ \
s=1,\ldots,m,
\eeq

In dimension $D$ (with gravitational constant $\kappa^2$),
the set of Einstein equations \linebreak
$R^M_N-R\delta^M_N/2=\kappa^2T^M_N$
can be written as
$R^M_N=\kappa^2[T^M_N-T\delta^M_N/(D-2)]$. Furthermore,
like the multidimensional geometry itself, these equations
decompose blockwise to
$R^{0}_{0}-R/2=\kappa^{2}T^{0}_{0}$ and
$R^{m_{i}}_{n_{i}}=\kappa^{2}[T^{m_{i}}_{n_{i}}-
T\delta^{m_{i}}_{n_{i}}/(D-2)]$. Using the previous formulas, we obtain
\beq{Einstein1}
\frac{1}{2}\sum_{i,j=1}^{n}G_{ij}\dot{x}^{i}\dot{x}^{j}+ V=0,
\eeq
\bear{Einstein2}
\ddot{x}^{i} + \dot{x}^{i}(\dot{\gamma_0}-\dot{\gamma})&=&
- \kappa^{2} \sum_{s=1}^{m}A^{(s)}
\left(
h^{(s)}_{i}(x) - \frac{\sum_{k=1}^{n}d_{k}h^{(s)}_{k}(x)}{D-2}
\right)\nonumber \\
&\times &
\exp\left[\Phi^{(s)}(x)
-2(\gamma-\gamma_{0})\right].
\end{eqnarray}
Here,
\beq{mini}
G_{ij}=d_{i}\delta_{ij}-d_{i}d_{j}
\eeq
are the components of the minisuperspace metric,
\beq{potential}
V=
\kappa^2\sum_{s=1}^{m}A^{(s)}
\exp\left[\Phi^{(s)}(x)
-2(\gamma-\gamma_0)\right].
\eeq
The integrals \rf{integrals} are used to replace the densities
$\rho^{(s)}$ in \rf{Einstein1}, \rf{Einstein2} by  the
functions $x^i(t)$.

It is not difficult to verify that after the gauge fixing
$\gamma=F(x^1, \ldots, x^n)$ the Einstein equations  \rf{Einstein2}
may be considered
as the Lagrange-Euler equations obtained from the Lagrangian
\beq{lagrangian}
L={\rm e}^{\gamma_0-\gamma}\left(
\frac{1}{2}\sum_{i,j=1}^{n}G_{ij}\dot{x}^{i}\dot{x}^{j}-V
\right)
\eeq
under the zero-energy constraint \rf{Einstein1}.

Further we develope an integration procedure which is based on the
$n$-dimensional Minkowsky-like geometry. Let $\R^n$ be the real
vector space and ${\bf e}_1,\ldots,{\bf e}_n$ be the canonical
basis in $\R^n$ (i.e. ${\bf e} _1= (1,0,\ldots,0)$ etc). Let us
define a symmetrical bilinear form $\langle,\rangle$ on $\R^{n}$
by \beq{product} \langle{\bf e}_i,{\bf e}_j\rangle=\delta_{ij}d_j
- d_i d_j \equiv G_{ij}. \eeq The form is nondegenerate and
inverse matrix to $(G_{ij})$ has  components \beq{inverse}
G^{ij} = \frac{\delta^{ij}}{d_{i}}+\frac{1}{2-D}.
\end{equation}
This form  $<.,.>$ endows the space $\R^n$ with a metric which
signature is $(-, +, ..., +)$ \cite{94}.
 With this in mind, a vector ${\bf
y}\in\R^n$ is timelike, spacelike or isotropic, if
$\langle{\bf y},{\bf y}\rangle$ takes negative, positive or null
values respectively and two vectors ${\bf y}$ and ${\bf z}$ are
orthogonal if $\langle{\bf y},{\bf z}\rangle=0$.

Hereafter, we use the following vectors
\bear{x}
{\bf x}&=&x^1(t){\bf e}_1+\ldots+x^n(t){\bf e}_n,\\
\label{u}
{\bf u}&=&u^1{\bf e}_1+\ldots+u^n{\bf e}_n,\
u^i=\frac{-1}{D-2},\ u_i=d_i,
\ear
where  covariant coordinates
$u_i$ of the vector ${\bf u}$ are introduced by the usual way.
Moreover, we obtain
\begin{eqnarray*}
\sum_{i,j=1}^{n}G_{ij}\dot{x}^{i}\dot{x}^{j}=
\langle\dot{\bf x},\dot{\bf x}\rangle=
\sum_{i=1}^n d_i(\dot{x}^i)^2 -\dot{\gamma}_0^2,\\
\langle{\bf u},{\bf x}\rangle=\gamma_0,
\ \langle{\bf u},{\bf u}\rangle=-\frac{D-1}{D-2}.
\end{eqnarray*}
\section{Exact solutions}
Now we suppose that all components are isotropic  fluids, i.e.
pressures $p^{(s)}_1,\ldots,p^{(s)}_n$ in the factor spaces
$M_1,\ldots, M_n$ are equivalent for each fluid
component. From the mathematical point of view it means that
\beq{FFF}
\Phi^{(s)}(x)=F^{(s)}(\gamma_0),\ \ s=1,\ldots,m,
\eeq where
$F^{(s)}$ is an arbitrary smooth function on $\R$ ($\gamma_0$ is
defined by \rf{gamma0}) and
 variable barotropic parameters
obtained from \rf{parameter} have the form
 $h^{(s)}_i(x)={\rm d}F^{(s)}(\gamma_0)/{\rm d}\gamma_0$ for all $i=1\ldots,n$.

We use the orthogonal basis
\beq{basis}
\frac{{\bf u}}{\langle {\bf u},{\bf u}\rangle},
{\bf f}_2,\ldots,{\bf f}_n\in\R^n,
\eeq
where the vector ${\bf u}$ was introduced by equation \rf{u}.
The orthogonality property reads
\beq{orthogonality}
\langle{\bf u},{\bf f}_j\rangle=0,\ \langle{\bf
f}_j,{\bf f}_k\rangle=\delta_{jk},\quad (j,k=2,\ldots,n).
\eeq
Let us note that  basis vectors ${\bf f}_2,\ldots,{\bf f}_n$
are space-like, since they are orthogonal to the time-like vector
${\bf u}$.  The vector ${\bf x}\in\R^n$ decomposes as follows
\beq{decomposition}
{\bf x}=
\gamma_0\frac{{\bf u}}{\langle{\bf u},{\bf u}\rangle}
+\sum_{j=2}^n\langle {\bf x},{\bf f}_j\rangle{\bf f}_j.
\eeq
We need to obtain the coordinates $\gamma_0,\langle {\bf x},{\bf f}_1\rangle,
\dots,\langle {\bf x},{\bf f}_n\rangle$
as functions of the time $t$.

Using the time gauge of the type
\beq{gauge}
\gamma=f(\gamma_0),
\eeq
we come to the Lagrangian \rf{lagrangian} in the terms of the coordinates
in such basis as follows
\beq{lagrangian1}
L=\frac{1}{2}
{\rm e}^{\gamma_0-f(\gamma_0)}
\left(
\frac{\dot\gamma_0^2}{\langle{\bf u},{\bf u}\rangle}
+
\sum_{j=2}^n\langle\dot{\bf x},{\bf f}_j\rangle^2
\right)
-
\kappa^2\sum_{s=1}^{m}A^{(s)}
{\rm e}^{f(\gamma_0)-\gamma_0+F^{(s)}(\gamma_0)}.
\eeq
The Lagrange-Euler equations for the coordinates
$\langle {\bf x},{\bf f}_1\rangle,
\dots,\langle {\bf x},{\bf f}_n\rangle$
\beq{LE}
\frac{\rm d}{{\rm d}t}
\left[
{\rm e}^{\gamma_0-f(\gamma_0)}
\langle\dot{\bf x},{\bf f}_j\rangle
\right]=0
\eeq
give immediatly the following integrals of motion
\beq{aj}
{\rm e}^{\gamma_0-f(\gamma_0)}
\langle\dot{\bf x},{\bf f}_j\rangle=a^j,\ \ j=2,\ldots,n,
\eeq
where $a^j$ is an arbitrary constant. Using \rf{aj}
we may present  the zero-energy constraint in the form
\beq{ze}
\dot\gamma_0^2 +
\langle{\bf u},{\bf u}\rangle
{\rm e}^{2[f(\gamma_0)-\gamma_0]}
\left(
\sum_{j=2}^n \left( a^j \right)^2 +
2\kappa^2
\sum_{s=1}^{m}A^{(s)}
{\rm e}^{F^{(s)}(\gamma_0)}
\right)=0.
\eeq
The last equation admits
obtaining of the unknown function $\gamma_0$
in quadratures for arbitrary functions
$f(\gamma_0),F^{(1)}(\gamma_0),\ldots,F^{(m)}(\gamma_0)$,
then the model is integrable.

To present  exact solutions in the Kasner-like form we
introduce the following vector
\beq{s}
{\bf s}=\sum_{j=2}^n a^j{\bf f}_j
\equiv
\sum_{i=1}^n s^i{\bf e}_i.
\eeq
Owing to the orthogonality property given by equation
\rf{orthogonality}
the coordinates  $s^i$ of this vector in the canonical
basis ${\bf e}_1,\ldots,{\bf e}_n$ satisfy the following constraints
\beq{constraints}
\langle{\bf s},{\bf s}\rangle=
\sum_{i,j=1}^n G_{ij} s^i s^j=
\sum_{j=2}^n \left( a^j \right)^2,\ \
\langle{\bf s},{\bf u}\rangle=
\sum_{i=1}^n d_i s^i=0
\eeq

Then the vector $\bf x$ we need to find may be presented as
\beq{decomposition2}
{\bf x}=
\gamma_0\frac{{\bf u}}{\langle{\bf u},{\bf u}\rangle}
+
\left[
\frac{{\rm sgn}(\dot\gamma_0)}
{\sqrt{-\langle{\bf u},{\bf u}\rangle}}
\int
\left(
\langle{\bf s},{\bf s}\rangle+
2\kappa^2\sum_{s=1}^{m}A^{(s)}
{\rm e}^{F^{(s)}(\gamma_0)}
\right)^{-1/2}
{\rm d}\gamma_0
\right]
{\bf s}.
\eeq

Here we
remind the multi-dimensional generalization of the well-known
Kasner solution \cite{92}.
It reads (for the synchronous time
$t_s$) as follows
\beq{Kasner} {\rm d}s^2=-{\rm
d}t^2_s+\sum_{i=1}^{n}A_it_s^{2\eps^i}{\rm d}s^2_i.
\eeq
Such a metric
describes the evolution of the vacuum model under consideration.
Kasner parameters $\eps^i$ satisfy the constraints
\beq{Kasnerconstraints}
\sum_{i=1}^nd_i\eps^i=1,\quad \sum_{i=1}^nd_i\left(\eps^i\right)^2=1.
\eeq
One may easily verify that the introduced parameters $s^i$ and
the Kasner parameters $\eps^i$ are connected by
\beq{connection}
\frac{{\rm sgn}(\dot\gamma_0)}
{\sqrt{-\langle{\bf u},{\bf u}\rangle}}
\cdot s^i=
\langle{\bf s},{\bf s}\rangle
\left(
\eps^i-\frac{u^i}{\langle{\bf u},{\bf u}\rangle}
\right),\ \
i=1,\ldots,n.
\eeq

Finally, we present the logarithms of  scale factors
in the Kasner-like form
\beq{final}
x^i=
\gamma_0\frac{u^i}{\langle{\bf u},{\bf u}\rangle}
+
\sqrt{\langle{\bf s},{\bf s}\rangle}
\int
\left(
\langle{\bf s},{\bf s}\rangle+
2\kappa^2\sum_{s=1}^{m}A^{(s)}
{\rm e}^{F^{(s)}(\gamma_0)}
\right)^{-1/2}
{\rm d}\gamma_0
\cdot
\left(
\eps^i-\frac{u^i}{\langle{\bf u},{\bf u}\rangle}
\right).
\eeq
We recall that coordinates of the vector $\bf u$ are defined by
\rf{u}. The function $\gamma_0(t)$ is determined by \rf{ze}.
If we use it as the time coordinate the exact solution
for the metric \rf{metr} looks as follows
\beq{metric}
{\rm d}s^2=
-\frac{D-2}{D-1}\cdot
\frac{{\rm e}^{2\gamma_0}}
{\langle{\bf s},{\bf s}\rangle+
2\kappa^2\sum_{s=1}^{m}A^{(s)}
{\rm e}^{F^{(s)}(\gamma_0)}}
\cdot
{\rm d}\gamma_0^2
+
\sum_{i=1}^{n}\exp[2x^i]{\rm d}s_i^2.
\eeq

\section{Examples}
\subsection{4-dimensional model}
Now we consider a special case with $D=4$ manifold
\begin{eqnarray*}
\M^4=\R\times M^1_1\times M^1_2\times M^1_3
\end{eqnarray*}
and a single perfect fluid. The dominant energy condition, applied
to the energy-momentum tensor \rf{tensor} reduces to the following
inequalities for the fluid variable barotropic parameter
$h=F'(\gamma_0)$
\begin{eqnarray*}
0\leq F'(\gamma_0)\leq 2.
\end{eqnarray*}
The exact solution for this special model looks as follows
\bear{x}
x^i=
\frac{\gamma_0}{3}
+
\sqrt{\langle{\bf s},{\bf s}\rangle}
\int
\left(
\langle{\bf s},{\bf s}\rangle+
2\kappa^2 A{\rm e}^{F(\gamma_0)}
\right)^{-1/2}
{\rm d}\gamma_0
\cdot
\left(
\eps^i-\frac{1}{3}
\right),\\
\label{rho}
\rho(t)=A
\exp\left[F(\gamma_0(t))-2\gamma_{0} (x)\right],
\ p(t)=\left[ 1-F'(\gamma_0(t))\right]\rho(t),
\ear
where the function $\gamma_0(t)=x^1(t)+x^2(t)+x^3(t)$ is the
solution to the separable ordinary
differential equation
\bear{ode}
\frac
{ \exp[\gamma_0-f(\gamma_0)] {\rm d} \gamma_0 }
{ \sqrt{ \langle{\bf s},{\bf s}\rangle + 2\kappa^2 A{\rm e}^{F(\gamma_0)} }  }
=
\pm \sqrt{ \frac{2}{3} }{\rm d}t.
\ear
We recall that  function $f(\gamma_0)$ defines a time gauge ($f(\gamma_0)\equiv 0$
corresponds to the synchronous time). The Kasner parameters $\eps^i$ obey the relations
\rf{Kasnerconstraints} with $d_1=d_2=d_3=1$. The nonnegative parameter $\langle{\bf s},{\bf s}\rangle$
is arbitrary. When $\langle{\bf s},{\bf s}\rangle=0$ one gets an isotropic solution, otherwise
the solution describes an anisotropic behaviour.

Let us analyze the behaviour of the exact solution assuming that the function $F(\gamma_0)$
is monotonically increasing with $F'(\gamma_0)>0$. We choose the
 following time gauge
 \begin{eqnarray*}
 {\rm e}^{ \gamma_0 - f(\gamma_0) }=
 F'(\gamma_0){\rm e}^{F(\gamma_0)}>0.
 \end{eqnarray*}
Then the equation \rf{ode} immediately  gives
\bear{odeint}
{\rm e}^{F(\gamma_0)}=
\frac{1}{3}A\kappa^2(t-t_0)\cdot
\left( t-t_0+\frac{\sqrt{6\langle{\bf s},{\bf s}\rangle}}{A\kappa^2}\right).
\ear
It follows  from \rf{odeint} that $F(\gamma_0)\to +\infty$ and
$\gamma_0\to +\infty$ as $t\to +\infty$. Then the equation \rf{x} presented
in the form
\bear{dx}
\frac{ {\rm d}x^i }{ {\rm d}\gamma_0 }=
\frac{1}{3} +
\langle{\bf s},{\bf s}\rangle
\frac
{ \eps^i - 1/3 }
{ \sqrt{\langle{\bf s},{\bf s}\rangle  +  2\kappa^2 A{\rm e}^{F(\gamma_0)} }}
\ear
shows an isotropic expansion of the universe with $x^i=\gamma_0/3 +$const, $i=1,2,3$, at late times.

At earlier times as $t\to t_0$ one gets
$ {\rm e}^{F(\gamma_0)}\to +0$ from the equation \rf{odeint}. Then the equations
\rf{ode},\rf{dx} lead to $\exp[x^i]\sim (t_s-t_{s0})^{\eps^i}$ in the main order as
the synchronous time $t_s\to t_{s0}+0$. This shows the Kasner-like behavior
near the initial singularity.

Hereafter we consider this single perfect fluid with  the variable barotropic equation of state
as an associated description of two different perfect fluids: normal matter with positive energy density
$\rho_m$ and nonnegative pressure $p_m=(1-h_m)\rho_m$ and quintessence with positive energy density $\rho_Q$
and negative pressure $p_Q=(1-h_Q)\rho_Q$. The  description implies
\bear{ass}
\rho=\rho_m+\rho_Q,\ \ p=p_m+p_Q.
\ear
If barotropic parameters $h_m,h_Q$ and $h=F'(\gamma_0)$ are specified, we get from the equations
\rf{rho},\rf{ass}
\begin{eqnarray*}
\rho_m=A\cdot\frac{ h_Q-F'(\gamma_0)}{h_Q-h_m}\cdot\exp\left[F(\gamma_0)-2\gamma_0\right],\\
\rho_Q=A\cdot\frac{F'(\gamma_0)-h_m}{h_Q-h_m}\cdot\exp\left[F(\gamma_0)-2\gamma_0\right].
\end{eqnarray*}
Then, the energy densities ratio looks as
\begin{eqnarray*}
\alpha=\frac{\rho_Q}{\rho_m}=\frac{F'(\gamma_0)-h_m}{ h_Q-F'(\gamma_0)}.
\end{eqnarray*}
In what follows we suppose that the barotropic parameters $h_m$ and $h_Q$ are constant and such that
\begin{eqnarray*}
0\leq h_m\leq 1,\ 1<h_Q\leq 2.
\end{eqnarray*}
Then the barotropic parameter of the  fluid
\begin{eqnarray*}
h=F'(\gamma_0)=h_Q-\frac{h_Q-h_m}{\alpha(\gamma_0)+1}
\end{eqnarray*}
is positive and the energy densities ratio $\alpha$ is a function of $\gamma_0$.
As we have already shown such model describes an isotropic expansion of the universe
at late times. If we suppose that the model is asymptotically coherent, i.e.
the energy densities ratio is asymptotically constant
\begin{eqnarray*}
\lim_{\gamma_0\to +\infty} \alpha(\gamma_0)=\alpha_0={\rm const},
\end{eqnarray*}
the scale factor and the density are in the main order
\begin{eqnarray*}
{\rm e}^{x^i}\sim t_s^{\frac{2}{3}\cdot\left(2-h_Q+\frac{h_Q-h_m}{\alpha_0+1}\right)^{-1}},
\ \rho\sim t_s^{-2}\ {\rm as}\ t_s\to +\infty,
\end{eqnarray*}
where $t_s$ is the synchronous time.
This is the late times isotropic asymptotic of the solution.
Further, if the following condition holds
\begin{eqnarray*}
\frac{h_Q-h_m}{\alpha_0+1}<h_Q - \frac{4}{3},
\end{eqnarray*}
the exact solution describes the power law isotropic
accelerated expansion at late times. If the normal matter is dust
($h_m=1$) and $\alpha_0$ according to the observations is equal to
$7/3$, then we obtain $h_Q>31/21$ or $p_Q<-10/21\cdot\rho_Q$.

\subsection{Multidimensional model}

Now we study the behaviour of the exact solution for the manifold
\begin{eqnarray*}
 \M=\R\times
M^3_1\times M_2^{d_2}\times \ldots \times M_n^{d_n},
\end{eqnarray*}
of the dimension $D=4+d$, where $M_1^3$ is the external flat
$3$-dimensional space and
\linebreak
$M_2^{d_2},\ldots, M_n^{d_n}$ are  internal spaces.
Let us admit that on some early stage of the evolution
 the functions $F^{(s)}(\gamma_0)$, $s=1,\ldots,m$, determining the barotropic
 parameters
$h^{(s)}_i(x)={\rm d}F^{(s)}(\gamma_0)/{\rm d}\gamma_0$, $s=1,\ldots,m$ and
$i=1,\ldots,n$, of the fluid components, obey the relation
\bear{rel}
\sum_{s=1}^{m}A^{(s)}
{\rm e}^{F^{(s)}(\gamma_0)}=\frac{A}{2\kappa^2}={\rm const}>0.
\ear
In  terms of  densities $\rho^{(s)}$, $s=1,\ldots,m$, the formula \rf{rel}
looks as follows
\begin{eqnarray*}
\sum_{s=1}^m \rho^{(s)}=\frac{A}{2\kappa^2}{\rm e}^{-2\gamma_0}.
\end{eqnarray*}
From the physical point of view this relation means that the
multicomponent perfect fluid appears as a whole as the stiff matter.
In this case the exact solution for the scale factors is
\bear{msf}
{\rm e}^{x^i(t_s)}=a_0^i\cdot(t_s-t_{s0})^{\tilde{\eps}^i},
\ear
where
\begin{eqnarray*}
\tilde{\eps}^i=
\frac{1}{D-1}+
\left( \eps^i-\frac{1}{D-1}\right)
\left(1+\frac{A}{\langle{\bf s},{\bf s}\rangle}\right)^{-1/2}.
\end{eqnarray*}

Using the Kasner-like constraints \rf{Kasnerconstraints} one easily obtains
the similar constraints for the parameters $\tilde{\eps}^i$
\beq{Kasnerlc}
\sum_{i=1}^nd_i\tilde{\eps}^i=1,
\quad \sum_{i=1}^nd_i\left(\tilde{\eps}^i\right)^2=
\frac{1}{D-1}+\frac{D-2}{D-1}\cdot
\left(1+\frac{A}{\langle{\bf s},{\bf s}\rangle}\right)^{-1}\equiv \delta.
\eeq
It is readily seen from \rf{Kasnerlc} that if the parameter
$A/\langle{\bf s},{\bf s}\rangle$ is small, then $\delta\approx 1$
and, consequently, $\eps^i\approx\tilde{\eps}^i$.
Then the model describes the Kasner-like behavior
of the multidimensional model
 with the expansion of one
part of the spaces $M_1^3,M_2^{d_2},\ldots,M_n^{d_n}$ and the contraction
of another part. From the physical viewpoint the behavior with expansion
of the external space $M_1^3$ and the contraction of internal space
(or spaces) is of the most interest. Let us analyze this solution
for the space-time manifold $\M^{4+d}=\R\times M_1^3\times M_2^d$, i.e. when
there is only one $d$-dimensional internal space $M_2^d$. In this case the solution
to the set of equations \rf{Kasnerlc} reads
\bear{tilde}
\tilde{\eps}^1=
\frac{1}{d+3}
\left(
1 \pm \frac{d}{3}\cdot\sqrt{3\cdot\frac{1+2/d}{1+A/\langle{\bf s},{\bf s}\rangle}}
\right),\
\tilde{\eps}^2=
\frac{1}{d+3}
\left(
1 \mp \sqrt{3\cdot\frac{1+2/d}{1+A/\langle{\bf s},{\bf s}\rangle}}.
\right)
\ear
We notice that  within the model the function
\begin{eqnarray*}
a_1^{(JBD)}={\rm e}^{x^1}={\rm const}\cdot
(t_s-t_{s0})^{\tilde{\eps}^1}
\end{eqnarray*}
is the scale factor of the external space with respect to the Jordan-Brance-Dicke frame
(see, for instance, \cite{GZ},\cite{BM},\cite{B}).
With respect to the Einstein frame the scale factor of the external space reads
\begin{eqnarray*}
a_1^{(E)}=
a_1^{(JBD)}\cdot {\rm e}^{d\cdot x^2/2}=
{\rm const}\cdot(t_s-t_{s0})^{\hat{\eps}^1},
\end{eqnarray*}
where
\bear{tildetilde}
\hat{\eps}^1=\tilde{\eps}^1+d\cdot\tilde{\eps}^2/2=
\frac{1}{d+3}
\left(
\frac{d+2}{2}
 \mp \frac{d}{6}
 \cdot
 \sqrt{3\cdot\frac{1+2/d}{1+A/\langle{\bf s},{\bf s}\rangle}}
\right)<\tilde{\eps}^1.
\ear
In order to obtain the contraction of the internal space (i.e. $\tilde{\eps}^2<0$)
we take the upper signs in the formulas \rf{tilde},\rf{tildetilde}
with the following obvious condition
\begin{eqnarray*}
\sqrt{3\cdot\frac{1+2/d}{1+A/\langle{\bf s},{\bf s}\rangle}}>1.
\end{eqnarray*}
Then $ A/{\langle{\bf s},{\bf s}\rangle} \in (0,2+6/d)$.
Using the last one easily gets  the following inequalities
\begin{eqnarray*}
\frac{1}{3}< \tilde{\eps}^1 <
\frac{1+\frac{d}{3}\cdot\sqrt{3(1+2/d)}}{d+3},\
\frac{d+2-\frac{d}{3}\cdot\sqrt{3(1+2/d)}}{d+3}<
\hat{\eps}^1 <\frac{1}{3}.
\end{eqnarray*}
The analysis
shows that if the solution describes  contraction of the internal space
then the external space in this solution may expand only with deceleration.

\begin{center}
{\bf Acknowledgments}
\end{center}

This work was supported in part by the Russian Foundation for Basic Research
(RFFI-04-02-16370-a) and CNRS.\\
V.N.M. is grateful to Prof. J.-M. Alimi for the hospitality
during his stay at LUTH, Observatory Paris-Meudon, France, in January, 2004.


\begin{thebibliography}{99}

\bibitem{Mel2}
Melnikov V.N. (1993) {\it Multidimensional Classical and Quantum Cosmology
and Gravitation. Exact Solutions and Variations of Constants},
CBPF-NF-051/93, Rio de Janeiro;\\
Melnikov, V.N. (1994) in: {\it Cosmology and
Gravitation}, ed. M. Novello, Editions Frontieres, Singapore, 1994, p. 147.

\bibitem{Mel}
Melnikov V.N. (1995) {\it Multidimensional Cosmology and  Gravitation},
CBPF-MO-002/95, Rio de Janeiro; \\
Melnikov V.N. (1996) In: {\it Cosmology and Gravitation. II}, ed. M.
Novello, Editions Frontieres, Singapore, p. 465.

\bibitem{Mel02}
Melnikov V.N. (2002)
 {\it Exact Solutions in Multidimensional Gravity and Cosmology III.} CBPF-MO-03/02,
 Rio de Janeiro, 297 pp.


\bibitem{3}
Staniukovich, K.P. and Melnikov, V.N. (1983) {\it Hydrodynamics, Fields and
Constants in the Theory of Gravitation}, Energoatomizdat, Moscow
(in Russian).\\
Melnikov, V.N. (2002)
{\it Fields and
Constants in the Theory of Gravitation}. CBPF-MO-02/02,
 Rio de Janeiro.

\bibitem{4}
Melnikov V.N. (1994) {\it Int. J. Theor. Phys.}  {\bf 33}, 1569.

\bibitem{5}
de Sabbata V., Melnikov V.N. and  Pronin P.I. (1992)
{\it Prog. Theor. Phys.} {\bf 88}, 623.

\bibitem{6}
Melnikov V.N. (1988) In: {\it Gravitational Measurements, Fundamental
Metrology and Constants}, eds. V. de Sabbata and V.N. Melnikov, Kluwer
Academic Publ., Dordtrecht, p. 283.

\bibitem{a1}
Gavrilov V.R. and Melnikov V.N.
Integration of $D$-dimensional Cosmological Models
with Two Factor-spaces by Reduction to the Generalized    Emden-Fowler Equation (1998)
{\it Teor. Mat. Fiz.} {\bf 114}, N 3, 454.

\bibitem{GMX}
Gavrilov V.R. and Melnikov V.N.
D-dimensional Integrable 2-component Viscous Cosmology
(2001) {\it Gavit. Cosmol.} {\bf 7}, N 4(28), 301.

\bibitem{Gavrilov}
Gavrilov V.R., Melnikov V.N. and Triay R.
Exact Solutions in Multidimensional Cosmology with Bulk and Shear Viscosity
(1997)
{\it Class. Quantum Grav.\/} {\bf 14}, 2203.

\bibitem{a2}
Bleyer U., Ivashchuk V.D., Melnikov V.N. and Zhuk A.I.
Multidimensional classical and quantum wormholes in theories
    with cosmological constant (1994) {\it Nucl.Phys.} {\bf B429}, 77.

\bibitem{a3}
Gavrilov V.R., Ivashchuk V.D. and Melnikov V.N.
 Multidimensional Integrable Vacuum Cosmology with Two Curvatures
  (1996) {\it Class.Quant.Grav.} {\bf 13},  3039.

\bibitem{a4}
Grebeniuk M.A., Ivashchuk V.D. and Melnikov V.N.
Multidimensional Cosmology for Intersecting p-Branes with Static Internal Spaces
 (1998) {\it Grav. Cosm.} {\bf 4}, N 2(14), 145.

\bibitem{a5}
Melnikov V.N.
Time Variations of G in Different Models (2002)
{\it Int.J.Mod.Phys.A} {\bf 17}, 4325.

\bibitem{a6}
Dehnen H., Gavrilov V.R. and Melnikov V.N.
General solutions for a flat Friedmann Universe Filled with  a Perfect
Fluid and a
 Scalar Field with an Exponential Potential
(2003) {\it Grav. Cosm.} {\bf 8}, N 4(32), 189.

\bibitem{a7}
Gavrilov V.R. and Melnikov V.N.
 2-Component Cosmological Models with Perfect Fluid and Scalar Fields. In:
 {\it Proceedings of the 18th Course of the School on
 Cosmology and Gravitation: The Gravitational Constant. Generalized Gravitational Theories
 and Experiments}
 (30 April-10 May 2003, Erice). Ed. by G. T. Gillies, V. N. Melnikov and V. de Sabbata,
 (Kluwer Acad.Publ.), (in print) 2003.

\bibitem{a8}
 Ivashchuk V.D., Selivanov A.B. and Melnikov V.N.
Cosmological solutions in multidimensional model with multiple exponential potential
  {\it JHEP} 0309 (2003) 059.


\bibitem{92}
Ivashchuk V.D.  (1992) {\it Phys. Lett.} {\bf A 170}, 16.

\bibitem{94}
Ivashchuk V.D. and Melnikov V.N. (1994) {\it Int. J. Mod. Phys.} {\bf D3},
795.

\bibitem{95}
Gavrilov V.R., Ivashchuk V.D. and Melnikov V.N. (1995) {\it J. Math. Phys.}
{\bf 36}, 5829.

\bibitem{96} Gavrilov V.R., Ivashchuk V.D. and Melnikov V.N. (1996)
{\it Class. Quantum Grav.} {\bf 13}, 3039.

\bibitem{GZ}
Gunther U. and Zhuk A.I. (1997) {\it Phys. Rev.} {\bf D56}, 6391.

\bibitem{BM}
Bronnikov K.A. and Melnikov V.N.
On observational predictions from multidimensional gravity (2001)
{\it Gen. Rel. Grav.} {\bf 33} 1549.


\bibitem{B}
Bronnikov K.A. and Melnikov V.N. Conformal frames and D-dimensional gravity.
In: {\it Proceedings of the 18th Course of the School on
Cosmology and Gravitation: The Gravitational Constant. Generalized Gravitational Theories
and Experiments}
(30 April-10 May 2003, Erice). Ed. by G. T. Gillies, V. N. Melnikov and V. de Sabbata,
 luwer Acad.Publ.), (in print) 2003.
\end{thebibliography}
\end{document}